\documentclass[conference]{IEEEtran}
\IEEEoverridecommandlockouts

\usepackage{lscape}
\usepackage{rotating}
\usepackage{epstopdf}
\usepackage{etoolbox}
\usepackage{nomencl}
\usepackage{amsthm}
\usepackage{supertabular}
\usepackage{amsfonts}
\usepackage[dvips]{epsfig}
\usepackage{latexsym}
\usepackage{amssymb}
\usepackage{amsmath}
\usepackage{verbatim}
\usepackage{setspace}
\usepackage[ruled,linesnumbered,boxed,commentsnumbered]{algorithm2e}
\usepackage{bm}
\usepackage[colorlinks=true, linkcolor=black, citecolor=blue, urlcolor=blue]{hyperref}       
\usepackage{url}            
\usepackage{lipsum}		
\usepackage{cite}
\usepackage{amsmath,amssymb,amsfonts}
\usepackage{algorithmic}
\usepackage{graphicx}
\usepackage{textcomp}
\usepackage{xcolor}
\usepackage{hyperref}
\usepackage[figure]{hypcap}
\hypersetup{hypertex=true,
            colorlinks=true,
            linkcolor=blue,
            anchorcolor=blue,
            citecolor=blue}
\def\BibTeX{{\rm B\kern-.05em{\sc i\kern-.025em b}\kern-.08em
    T\kern-.1667em\lower.7ex\hbox{E}\kern-.125emX}}
\begin{document}

\title{Fairness-Aware Optimization of Vehicle-to-Vehicle Interaction for Smart EV Charging Coordination
\thanks{\dag These authors contributed equally.}
\thanks{*Corresponding author: Hao Wang.}
\thanks{This work has been supported in part by the Australian Research Council (ARC) Discovery Early Career Researcher Award (DECRA) under Grant DE230100046.}
}

\author{\IEEEauthorblockN{Aditya Khele$^\dag$}
\IEEEauthorblockA{\textit{Faculty of Information Technology} \\
\textit{Monash University}\\
Melbourne, Australia \\
kheleaditya@gmail.com}
\and
\IEEEauthorblockN{Canchen Jiang$^\dag$}
\IEEEauthorblockA{\textit{Department of Data Science and AI} \\
\textit{Monash University}\\
Melbourne, Australia \\
canchen.jiang@monash.edu}
\and
\IEEEauthorblockN{Hao Wang$^*$}
\IEEEauthorblockA{\textit{Department of Data Science and AI } \\
\textit{and Monash Energy Institute}\\
\textit{Monash University}\\
Melbourne, Australia \\
hao.wang2@monash.edu}
}

\maketitle

\begin{abstract}
As the number of electric vehicles (EVs) continues to grow, there is an increasing need for smart charging strategies. This paper exploits the vehicle-to-vehicle (V2V) concept to leverage EVs' diverse charging patterns and unlock the value of flexibility by enabling energy transfer among EVs. We formulate a cost minimization problem for an EV charging station to optimize the V2V schedule together with vehicle-to-grid (V2G), grid-to-vehicle (G2V) charging, as well as the use of renewable energy. When EVs perform V2V to transfer energy to charge EVs, the traditional cost-minimizing approach may overuse some EVs with lower costs to perform V2V. We address the fairness issue by developing fair V2V energy transfer strategies to avoid the excess discharge from individual EVs. We introduced three kinds of fairness metrics in the V2V optimization problem to demonstrate the fair energy transfers. In addition, we formulate baseline optimization problems without V2G or V2V and compare the results to prove the potency of V2V concept along with V2G method. The simulation results demonstrate that V2V can significantly reduce EV charging costs and highlight the trade-off between fairness enforcement and charging cost minimization.
\end{abstract}


\makenomenclature


\renewcommand\nomgroup[1]{%
  \item[\bfseries
  \ifstrequal{#1}{A}{Abbreviations}{%
  \ifstrequal{#1}{V}{Variables}{%
  \ifstrequal{#1}{P}{Parameters}{%
  \ifstrequal{#1}{N}{Number sets}{%
  \ifstrequal{#1}{I}{Indices}{}}}}}%
]}

\nomenclature[A]{\(\text{EV}\)}{\quad \quad Electric Vehicle}
\nomenclature[A]{\(\text{SOC}\)}{\quad \quad State Of Charge}
\nomenclature[A]{\(\text{PV}\)}{\quad \quad Photovoltaic}
\nomenclature[A]{\(\text{MIQP}\)}{\quad \quad Mixed-Integer Quadratic programming}
\nomenclature[A]{\(\text{AEMO}\)}{\quad \enspace \thinspace Australia Energy Market Operator}
\nomenclature[A]{\(\text{V2G}\)}{\quad \quad Vehicle-to-Grid}
\nomenclature[A]{\(\text{V2V}\)}{\quad \quad Vehicle-to-Vehicle}
\nomenclature[P]{\(\lambda^i_c\)}{Charging efficiency coefficient of the EV $i$}
\nomenclature[P]{\(\lambda^i_d\)}{Discharging efficiency coefficient of the EV $i$}
\nomenclature[P]{\(\alpha^i\)}{The amortized cost coefficient of the EV $i$ battery discharging}
\nomenclature[P]{\(G\)}{The maximal power that all EVs can purchase from grid}
\nomenclature[P]{\(R_{Renew}^t\)}{The maximal renewable energy supply per each EV at time $t$}
\nomenclature[P]{\(R^{u,t}\)}{The maximal PV energy that prosumer $u$ can utilize in time slot $t$}
\nomenclature[P]{\(\overline{P}_{ch}^{i,t}\)}{The maximum limit for charging for EV $i$ in time slot $t$}
\nomenclature[P]{\(\overline{P}_{dis}^{i,t}\)}{The maximum limit for discharging for EV $i$ in time slot $t$}
\nomenclature[P]{\(\overline{P}_{V2G}^{i,t}\)}{The maximum limit for EV $i$ to sell energy back to grid in time slot $t$}
\nomenclature[P]{\(\overline{P}_{V2V}^{i,j,t}\)}{The maximum limit for EV $i$ to transfer energy to EV $j$ in time slot $t$}
\nomenclature[P]{\({\phi}_G^t\)}{Price of electricity from the grid at time $t$}
\nomenclature[P]{\(\overline{\mathcal{Z}}\)}{The maximum limit for EV $i$ to discharge in time slot $t$}
\nomenclature[P]{\(\overline{\mathcal{Z}^c}\)}{The maximum limit for EV $i$ to discharge over the period $\mathcal{H}$}
\nomenclature[P]{\(\mathcal{D}_{i}^c\)}{The maximum limit for EV $i$ to discharge over the given threshold in period $\mathcal{H}$}

\nomenclature[P]{\(P_{target}^{i, t_{d}^{i}}\)}{Required energy for EV $i$ at it's departure time}

\nomenclature[V]{\(p_{ch}^{i,t}\)}{The amount of charge power in EV $i$ in $t$-th time slot}

\nomenclature[V]{\(p_{dis}^{i,t}\)}{The amount of discharge power of EV $i$ in $t$-th time slot}
\nomenclature[V]{\(p_{V2V}^{i,j,t}\)}{The amount of energy from EV $i$ to EV $j$ in $t$-th time slot}
\nomenclature[V]{\(p_{avail}^{i,t}\)}{The stored energy of EV $i$ in $t$-th time slot}
\nomenclature[V]{\(P_{Renew}^{i,t}\)}{Renewable energy usage of EV $i$ in the t-th time slot}
\nomenclature[V]{\(P_{grid}^{i,t}\)}{The amount of electricity that EV $i$ purchases from the grid in the $t$-th time slot}
\nomenclature[V]{\(P_{V2G}^{i,t}\)}{The V2G energy that EV $i$ sells to the grid at time $t$}
\nomenclature[V]{\(X_{t}^i\)}{Binary variable that limit EV charging and discharging simultaneously in EV $i$ in time slot $t$}
\nomenclature[V]{\(t^{i}_{a}\)}{Arrival time of EV $i$}
\nomenclature[V]{\(t^{i}_{d}\)}{Departure time of EV $i$}
\nomenclature[V]{\(\mathcal{Z}_{i}^{t}\)}{Excess amount of discharge than threshold for EV $i$ at time $t$ }
\nomenclature[N]{\(\mathcal{N}\)}{The set of EVs}
\nomenclature[V]{\(\theta\)}{Measure of threshold at time $t$ for each EV participating in the discharge}
\nomenclature[P]{\(C_{total}\)}{The effective cost of charging EVs parked in the considered parking space}
\nomenclature[N]{\(\mathcal{H}\)}{The set of time slots}
\nomenclature[I]{\(i\)}{The $i$-th EV}
\nomenclature[I]{\(j\)}{The $j$-th EV}
\nomenclature[I]{\(t\)}{The $t$-th time slot}

\printnomenclature

\section{Introduction}
The electric vehicle (EV) market is rapidly growing due to its potential as a promising technical alternative to fossil-fuel-powered automobiles in the fight against climate change \cite{ajanovic2015future}. The number of EVs on the road has increased by 60\% since 2018, and the reduced cost of high-performance battery technology further boosts their development, leading to a considerable impact on the electric grid \cite{abronzini2019cost}. Uncoordinated charging of a large number of EVs could stress the grid, leading to voltage issues and congestion. Therefore, the development of smart charging and vehicle-to-grid (V2G) technologies are essential for the effective energy management of EVs, enabling the seamless integration of large numbers of EVs without compromising grid stability and reliability.

Coordinated EV charging is a fundamental aspect of ensuring a sustainable and reliable power grid that can accommodate the growth of EVs and enable the transition to a cleaner energy future. Recently, various EV charging approaches using coordination, pricing, and incentives have been proposed, e.g., in \cite{he2012optimal,suyono2019optimal,RN105}. 
Organized discharging technologies, such as V2G, allow bidirectional energy exchange between electric vehicles and the power grid. It can aid the power grid in reducing the peak load \cite{liu2013opportunities,tan2016integration}. In V2G, a significant number of EVs are utilized as both load and energy storage to support the grid by storing energy during off-peak hours and selling it back to the grid during peak hours \cite{V2G2019}. However, the above studies only explored the interaction between EVs and the grid but neglected the potential value brought by interactions among EVs. \emph{This research aims to unlock the value of EV flexibility through vehicle-to-vehicle (V2V) energy transfer while taking EV fair contribution into account.}

\subsection{Existing Studies and Gap}
Several studies have focused on EV energy management which can help reduce the cost of charging EVs and reduce the energy load for the grid at peak hours.
A large-scale EV integration, also known as grid-to-vehicle (G2V), has been studied to determine its effects on the power system. The study in \cite{yi2020highly} demonstrated that controlled EV charging could lower the peak demand by distributing the majority of the peak load. Moreover, \cite{sarker2016co} showed the potential profitability from lowering energy peak load and customer costs together. 
A priority-based V2G peak load optimization approach explained in \cite{hashim2021priority}, using grid-connected EVs, reduces the disparity between peak demand and off-peak load. As charging stations heavily rely on the grid power to charge EVs, it is more environmentally friendly to integrate renewable energy sources with V2G energy transfer \cite{iacobucci2019optimization}. In addition to drawing local renewable energy to charge EVs, vehicle-to-vehicle (V2V), as an emerging technology, demonstrates that EVs may transfer their energy via bidirectional chargers in a charging station. V2V enables flexible power transfer between EVs, making EV charging coordination smarter and more efficient \cite{liu2013opportunities}. 

Vehicle-to-vehicle charging is a technology that allows electric vehicles to use their batteries to exchange energy with other vehicles. 
It is a relatively new technology that emerged because EV charging patterns are diverse, and some EVs connected to charging stations can share the energy to help each other and reduce the peak load of the charging station and the grid. In the last few years, a few aspects of V2V operations have been studied. A fuzzy energy management algorithm was developed in \cite{mohamed2013real} to reduce the overall daily cost of charging the EVs and mitigate the impact of EV charging on the main grid. 
A study in \cite{Kim2018} proposed a matching-based method to assign a charging station or an EV as a supplier to meet the charging demand of EVs, supported by V2V. 
The study in \cite{liu2022reservation} proposed a V2V charging strategy, which incorporates parking duration to facilitate V2V charging under time restraints. This research suggested that V2V can offer the charge services to grid-connected electric vehicles to offload the power system’s power demands when needed. However, discharging of the electric vehicles should be restricted and allow all EVs to benefit from respective discharging. 

EVs could perform a variety of tasks, such as V2G and V2V, all of which generate value streams. Similar to photovoltaic (PV) energy exporting, EVs can discharge their batteries to become energy suppliers. 
Energy customers used to be solely consumers, but they have evolved into prosumers with the capacity to generate surplus revenue. This changes the energy landscape and fair allocation of benefits has emerged as a problem. If it is not addressed, it may constitute a barrier to the adoption of EVs and PV systems. 
For example, in distribution networks with PV generation, unfair distribution of energy is a significant issue. Studies have addressed this unfair energy supply issue for PV curtailment. 
The study in \cite{poudel2022fairness} presented a fairness-based energy coordination strategy enabling customers to share the responsibility of voltage regulation in distribution systems. Fair PV curtailment was implemented in \cite{gebbran2021fair} utilizing various notions of fairness, resulting in varied curtailment patterns.
Similarly, EV discharging needs to be constrained to benefit all participating EVs and prevent the over-discharge of one or a few EVs, thus providing a fair chance to all EVs. 
Existing solutions to fair PV curtailment could not be directly applied in the relation to V2V fairness. As in the case of fairness in PV curtailment, energy flow is unidirectional whereas fairness in V2V technology energy flow is bidirectional. 
In this paper, along with different energy-exchanging strategies, we focus on reducing the biased discharging of electric vehicles with the flexibility to charge and discharge the vehicles over a period. 

\subsection{Main Work and Contributions}
In this paper, we study the fairness-aware optimization of V2V to facilitate more flexible EV charging to reduce energy costs while explicitly ensuring fair discharge of EVs. The EV charging is co-optimized with V2V, V2G, and local PV energy supply. 
The V2V technology leverages the EV interactions and benefits the system by transferring energy to EVs in need, thus reducing the overall energy purchase from the grid. To avoid excess and biased discharging from a particular group of EVs, we introduce the fairness concept to enforce fair discharging of EVs participating in V2G and V2V activities.
The key contributions of this paper are summarized as follows.
\begin{itemize}
  \item \textit{Optimizing V2V for Smart Charging}: We study smart charging techniques using V2V to leverage the EV flexibility and benefit the EV users and the grid. V2V leverages the diverse EV charging patterns and bidirectional chargers to fully unlock the value of EV flexibility.
  \item \textit{Fair Discharging and Impact}: We design three different fairness constraints to avoid excess discharging of individual EVs and maintain balanced energy transfer contributions. The designed fairness constraints enable an explicit analysis of the tradeoff between EV charging cost minimization and fairness enforcement. 
  \item \textit{Value of V2V in Residential and Shopping Center EV Charging Scenarios}: We demonstrate the value of V2V in a smart charging environment to balance the load and reduce the charging cost of EVs parked in residential and shopping center EV charging stations.
\end{itemize}
The remainder of this paper is organized as follows. Section \ref{sec:systemmodel} presents the system model. Section \ref{sec:fairness} presents three fairness constraints for EV discharging. Section \ref{sec:formulation} presents the joint optimization problem for EV charging, V2G, and V2V. It also includes baseline optimization problems for comparison. Section \ref{sec:simulation} evaluates our proposed fairness-aware V2V optimization. Section \ref{sec:conclusion} concludes our work.

\section{System Model}\label{sec:systemmodel}

\subsection{System Setup}
\begin{figure}[!th]
  \includegraphics[width=0.95\linewidth]{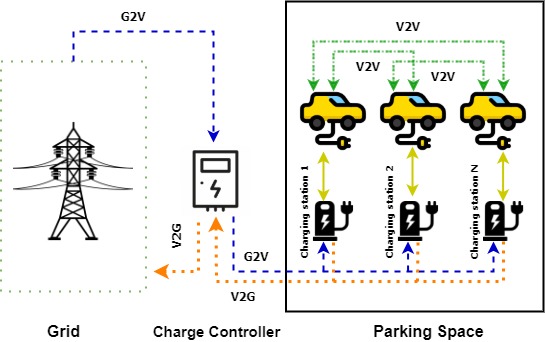}
  \caption{System Diagram of EV charging with V2G and V2V.}
  \label{fig:system}
\end{figure}

We consider a system with a large parking space, such as a residential premise and shopping center, having multiple EV chargers. As shown in Fig.~\ref{fig:system}, the green dot-dashed lines represent the energy transfer between EVs performing V2V, and the orange dotted lines signify V2G. The dashed blue lines denote the EV charging.  A controller is responsible for overall energy transactions with the grid and coordinating EV charging and discharging operations in the charging site. This problem aims to minimize the total cost for EVs and enable fair participation of EVs in energy transfer. This section presents the system model, including EV, energy supply, V2G, and V2V.

\subsection{System Model}

We denote a set of EVs as $\mathcal{N} = \{ 1,..., N \}$, where $N$ is the total number of EVs. We consider the operational horizon of a day, which is evenly divided into $24$ time slots, such that each time slot denotes one hour. We denote the operational horizon as $\mathcal{H} = \{ 1,..., H \}$, where $H = 24$ is the total number of time slots.

\subsubsection{EV battery model}
For EV $i \in \mathcal{N}$, the parking period is denoted as $[t^{i}_{a}, t^{i}_{d}]$ $\subseteq$ $H$. At the arrival time $t^{i}_{a}$, EV $i$ is connected to the charger. Similarly, $ t^{i}_{d}$ represents the departure time of EV $i$, when EV $i$ is plugged out. 
We denote $P_{avail}^{i,t}$ as the available energy stored in EV $i$ at time $t$, $P^i_{ch,t}$ as the charge energy, and $P^i_{dis,t}$ as the discharge energy of EV $i$ in time slot $t$. We let
$\lambda^i_c$ $\in$ [0,1] and $\lambda^i_d$ $\in$ (0,1] denote charging and discharging efficiency of EV $i$, respectively.
The battery dynamics of EV $i$ in time slot $t$ is presented as
\begin{equation}
p_{avail}^{i,t} = p_{avail}^{i,t-1} + \lambda_c p_{ch}^{i,t} - \frac{1}{\lambda_d}  p_{dis}^{i,t}. \label{cons:dynamics}
\end{equation} 

Since EVs cannot charge and discharge at the same time, we restrict the EV operations by introducing a binary variable denoted as $X_{t}^i \in \{0,1\}$, in which $X_{t}^i = 0$ represents charging, and $X_{t}^i = 1$ represents discharging.
The constraint for EV charging at  time $t \in [t^{i}_{a}, t^{i}_{d}]$ is
\begin{equation}
0 \leq p_{ch}^{i,t} \leq \overline{P}_{ch}^{i,t}(1 - X_{t}^i).
\end{equation}

Similarly, the constraint for EV discharging at time $t \in [t^{i}_{a}, t^{i}_{d}]$ is
\begin{equation}
0 \leq p_{dis}^{i,t} \leq \overline{P}_{dis}^{i,t} X_{t}^i,
\end{equation}
where $\overline{P}_{ch}^{i,t}$ and $\overline{P}_{dis}^{i,t}$ denote the upper bound for $P_{ch}^{i,t}$ and $P_{dis}^{i,t}$ in all time slots $t \in [t^{i}_{a}, t^{i}_{d}]$, respectively. When the value of $X_{t}^i$ is $1$, the upper bound for $p_{ch,t}^{i}$ becomes $0$, allowing EV $i$ to discharge. When the value of $X_{t}^i$ is $0$, the upper bound for $p_{dis}^{i,t}$ becomes 0, allowing EV $i$ to charge. When EVs are not present at the parking slot, the values of $p_{ch}^{i,t}$ and $p_{dis}^{i,t}$ are $0$.

The stored energy in EV $i$ is bounded by
\begin{equation}
\underline{P}_{avail}^{i,t} \leq p_{avail}^{i,t} \leq \overline{P}_{avail}^{i,t},
\end{equation}
where $\underline{P}_{avail}^{i,t}$ and $\overline{P}_{avail}^{i,t}$ are the lower and upper bounds for $p_{avail}^{i,t}$ in all time slots $t \in [t^{i}_{a}, t^{i}_{d}]$.

We assume that each EV provides information at arrival time about the amount of required energy to be charged by the departure time. We denote the required charge by $P_{target}^{i, t_{d}^{i}}$ for EV $i$. The constraint for required energy at departure time is
\begin{equation}
p_{avail}^{i, t_{d}^{i}} = P_{target}^{i, t_{d}^{i}}.
\end{equation}

\subsubsection{Energy supply model}
The energy supply can come from the grid $P_{grid}^{i,t}$ and the local renewable generation $P_{Renew}^{i,t}$, e.g., photovoltaic (PV) energy, for EV $i$ in time slot $t$. The constraints for energy supply are
\begin{align}
& 0 \leq \sum_{i \in \mathcal{N}}p_{grid}^{i,t} \leq G,\\
& 0 \leq \sum_{i \in \mathcal{N}}p_{Renew}^{i,t} \leq R_{Renew}^t,
\end{align}
where $G$ is the upper bound for grid power, and $R_{Renew}^t$ is the upper bound for renewable supply per each EV at time $t$.

\subsubsection{V2V and V2G energy transactions}
Energy transactions consist of V2G and V2V. 
For V2G, we denote $p_{V2G}^{i,t}$ as the amount of energy that EV $i$ performs V2G to sell to the grid. Through V2G, EV batteries deliver services to the electrical grid, assisting in the maintenance of high-quality and reliable electricity supply \cite{shirazi2015cost}. The constraint for $p_{V2G}^{i,t}$ is
\begin{equation}
0 \leq p_{V2G}^{i,t} \leq \overline{P}_{V2G}^{i,t},
\end{equation}
where $\overline{P}_{V2G}^{i,t}$ is the upper bound for the amount of V2G for EV $i$ at time $t$. 

Another energy transaction performed among EVs is V2V, and we denote $p_{V2V}^{i,j,t}$ as the energy transfer between EVs $i,j \in \mathcal{N}$ at time $t$. The constraint for V2V is
\begin{equation}
0 \leq p_{V2V}^{i,j,t} \leq \overline{P}_{V2V}^{i,j,t},
\end{equation}
where $\overline{P}_{V2V}^{i,j,t}$ is the upper bound for V2V transfer at time $t$. 

Note that, when EV $i$ transfers its energy to EV $j \in \mathcal{N} \backslash i$, we assume that the energy loss among EVs in the charging site is negligible, and thus EV $j$ receives the same amount of energy, which can be represented as 
\begin{equation}
p_{V2V}^{i,j,t} = -p_{V2V}^{j,i,t}.
\end{equation}

Here, $p_{V2V}^{i,j,t}$ denotes the charging of EV $i$ from EV $j$ and $-p_{V2V}^{j,i,t}$ denotes discharging of EV $j$ to charge EV $i$ at time $t$. 

The energy coming into the system and going out of the system should be balanced out. The total energy charge consists of energy supply from the grid and renewable and the energy received from V2V
\begin{equation}
p_{ch}^{i,t} = p_{grid}^{i,t} + p_{Renew}^{i,t} + p_{V2V}^{j,i,t}.
\end{equation}

We consider the EV-optimized scheduling charging and discharging depending upon the different values at different times. The discharging of the EV satisfies the following constraint
\begin{equation}
p_{dis}^{i,t} = p_{V2G}^{i,t} + p_{V2V}^{i,j,t}.
\label{cons:common_constraints}
\end{equation}

\section{Fairness Metrics and Constraints}\label{sec:fairness}
Fair discharging for V2G or V2V is necessary for participated EVs. For example, the V2V technology enables energy transfer among connected EVs, intending to offload significant charging load on the grid \cite{wang2014semi}. Similar to demand response and PV curtailment, V2V interaction can lead to biased discharging from specific EV groups and unfair contributions from EVs. Some EVs excessively contribute to the V2V energy transfer, causing a burden, but some do not get an equal opportunity to participate. 

Our work focuses on the fair discharging of EVs participating in V2G and V2V activities. We consider fairness constraints to prevent the over-discharging of parked EVs and provide an opportunity for all EVs to benefit from V2G and V2V technologies. In this paper, we consider and discuss two methods for introducing fairness. In the first approach, we define a strictly fixed limit for each EV participating in discharge activity. In this approach of fairness constraints with fixed limits, we further define two fairness constraints with fixed hard and soft limits separately. A more flexible EV discharging fairness solution is discussed in the second approach. The detailed explanation of these fairness constraints is as follows.

\subsection{Definition 1: Fairness constraints with fixed hard limit}
Firstly, we consider stricter constraints where the amount of discharge is restricted to the given limit for all EV $i$ $\in$ $\mathcal{N}$ over time $t$. We define the upper boundary for each discharging EV $i$ for time $t$ by $\overline{\mathcal{Z}}$. The upper limits will strictly restrict the excess discharging of the EVs and hence, allowing EVs to discharge fairly. We define this by
\begin{equation}
p_{dis}^{i,t} \leq \overline{\mathcal{Z}}, \: \: \forall \: t,
\label{cons:fair1}
\end{equation}

This constraint ensures fairness in discharging EVs while providing a fair chance to every EV at $t$ parked in the parking space to perform V2V and/or V2G.

\subsection{Definition 2: Fairness constraints with fixed soft limit}
The second considered fairness constraint with a fixed limit also restricts the discharge of EVs but the applied limit is for cumulative discharge by each EV over the horizon $\mathcal{H}$. Fairness constraints with a soft limit allow EVs to discharge freely till the total cumulative amount of discharge reach or is less than the defined limit. To define this, we consider $\overline{\mathcal{Z}^c}$ to represent the threshold for the total amount of discharged energy for each EV participating in the discharging activity over the time period $\mathcal{H}$. This can be represented by
\begin{equation}
\sum_{t \in \mathcal{H}} p_{dis}^{i,t} \leq \overline{\mathcal{Z}^c}.
\label{cons:fair2}
\end{equation}

Note that the thresholds $\mathcal{Z}$ and $\overline{\mathcal{Z}^c}$  need to be tuned for the problem, and we will present the effect of different thresholds in Section~\ref{sec:simulation}.

\subsection{Definition 3: Fairness constraints with budget}
We consider static constants\footnote{There are different definitions for fairness, such as pair-wise comparisons between EVs, leading to an exponential number of constraints. In this work, we choose computationally efficient yet effective constraints for fairness.} $\theta$ and $\mathcal{D}_{i}^c$  for fairness in the EV discharging activities. We define $\theta$ as the measure of a threshold at time $t$ for each EV participating in the discharge activity. To maintain the flexibility of the model, we allow vehicles to discharge over the threshold $\theta$ at time $t$ and consider minimizing the excess discharging by limiting the total excess discharge by $\mathcal{D}_{i}^c$. The excess amount of discharge of energy can be denoted by 
$(p_{dis}^{i,t} - \theta)^{+}.$ The amount of excess discharge can only be a positive entity for any EV $i$ at any time $t$. The fairness equation with budget $\mathcal{D}_{i}^c$ represented as
\begin{equation}
\sum_{t \in \mathcal{H}} \mathcal (p_{dis}^{i,t} - \theta)^{+} \leq \mathcal{D}_{i}^c.
\label{cons:fair3}
\end{equation}

The results from both approaches with different parameter settings in each of case will be discussed in Section~\ref{sec:simulation}.

\section{Optimization Problem Formulation}\label{sec:formulation}
After presenting the models for EV charging, V2G, V2V, and fairness, we formulate an optimization problem to jointly schedule EV charging, V2G, and V2V subject to fairness and other operational constraints for minimizing the total cost. In the following, we will present the costs and the revenue of the charging site. 

\subsection{Grid energy purchase cost}  
For the cost calculation of energy purchased from the grid, we consider a time-of-use pricing scheme \cite{suyono2019optimal}. The most important characteristic of time-of-use pricing is that it provides end-users with time-differentiated energy prices. We consider three predefined price levels per day for peak, off-peak, and shoulder periods. The energy purchase cost from the grid for EV $i$ is
\begin{equation*}
C_{grid}^{i} = \sum_{t \in \mathcal{H}} {\phi}_G^t p_G^{i,t}, 
\end{equation*}
where ${\phi}_G^t$ is the predefined time-of-use price.

\subsection{V2G revenue earned from the grid}   
We consider dynamic electricity prices from the Australian Energy Market Operator (AEMO) for V2G energy transactions. AEMO is responsible for the day-to-day operations of a variety of electricity and gas markets and information services, as well as strategic forecasting and planning \cite{aemo}. We are using half-hourly prices retrieved from AEMO.
Based on the retrieved prices, EVs can sell energy back through an aggregator or retailer to the grid and generate revenue. The revenue at time $t$ of EV $i$ is denoted as ${\phi}_{V2G}^t p_{V2G}^{i,t}$. Hence, the total revenue of V2G for EV $i$ is
\begin{equation*}
R_{V2G}^{i} = \sum_{t \in \mathcal{H}} \phi_{V2G}^t p_{V2G}^{i,t}.
\end{equation*}

\subsection{Battery degradation cost}
The discharging of the EV batteries leads to a progressive alteration in the physical structure of both the electrolyte and the electrode, resulting in battery degradation \cite{8949354}. We consider the battery degradation cost based on the amount of EV energy discharging while formulating the optimization problem. The cost of an EV battery degradation for EV $i$ is
\begin{equation*}
C_{battery}^{i} = \alpha^i \sum_{t \in \mathcal{H}} (p_{dis}^{i,t})^2,
\end{equation*}
where $\alpha^i$ is the cost coefficient of EV $i$'s battery.

\subsection{Joint optimization problem for EV charging, V2G, and V2V}
We formulate the optimization problem to jointly optimize the EV charging, V2G, and V2V schedules in the parking site. We denote $c_{total}$ as the total cost of charging all the EVs in the parking space till the specified individual target. The objective is to minimize the $c_{total}$ by scheduling charging and discharging of the EVs through variables including $p_{grid}^{i,t}$, $p_{V2G}^{i,t}$ and $p_{V2V}^{j,i,t}$. Hence, the optimization problem is formulated as
\begin{equation}
\begin{aligned}
& \min \quad && C_{total} = \sum_{i \in \mathcal{N}} (C_{grid}^{i}+ C_{battery}^{i}- C_{V2G}^{i})\\
& \text{subject to} && \text{Constraints}~ \eqref{cons:dynamics}-\eqref{cons:common_constraints}
\text{ \& Fairness Constraints},\label{obj:equation}
\end{aligned}
\end{equation}
where three types of fairness constraints are presented in Section~\ref{sec:fairness}. We include each type in the optimization problem to evaluate its effectiveness, separately.

In this formulated joint optimization problem, fairness constraint can be either one of the defined constraints \eqref{cons:fair1}, \eqref{cons:fair2} or \eqref{cons:fair3}. In the case of fairness constraint with a budget defined in \eqref{cons:fair3}, it's equivalent to solve the optimization problem \eqref{obj:equation} using the following constraints
\begin{align}
& \mathcal{Z}_{i}^{t} \geq p_{dis}^{i,t} - \theta,
\label{cons:fair3_eq1} \\
& \mathcal{Z}_{i}^{t} \geq 0, \label{cons:fair3_eq2} \\
& \sum_{t \in \mathcal{H}} \mathcal{Z}_{i}^{t} \leq \mathcal{D}_{i}^c,\label{cons:fair3_eq3}
\end{align}
where we denote excess amount of discharge for vehicle $i$ at time $t$ by $\mathcal{Z}_{i}^{t}$ and $\mathcal{Z}_{i}^{t}$ is non-negative entity for all $i$ at any time $t$.

The formulated optimization problem is a Mixed-Integer Quadratic programming (MIQP) problem. Furthermore, baseline optimization problems \eqref{obj:baseline1}, \eqref{obj:baseline2} are presented in Section~\ref{sec:baseline}. To solve the optimization problem \eqref{obj:equation} and baseline optimization problems \eqref{obj:baseline1} and \eqref{obj:baseline2}, we used Gurobi Solver.

\subsection{Baseline Optimization Problems}\label{sec:baseline}
We construct two baseline optimization problems to compare them to the proposed joint optimization problem and evaluate the benefit of V2V. First, we consider a baseline problem, which only optimizes EV charging with no V2G and V2V energy transactions. Second, we consider two baseline problems to jointly optimize EV charging with V2G.
\begin{itemize}
    \item For the first baseline problem, we only consider EV charging. Hence, we set $p_{V2G}^{i,t}=0$ and $p_{V2V}^{i,t}=0$, such that the EV charging uses grid power and local renewable, i.e., $p_{ch}^{i,t} = p_{grid}^{i,t} + p_{Renew}^{i,t}$, and there is no discharge. The objective function for baseline without considering V2G and V2V given by,
\begin{equation}
\begin{aligned}
& \min \quad && C_{total} = \sum_{i \in \mathcal{N}} (C_{grid}^{i}+ C_{battery}^{i})\\
& \text{subject to} && \text{Constraints}~ \eqref{cons:dynamics}-\eqref{cons:common_constraints} .
 \label{obj:baseline1}
\end{aligned}
\end{equation}

\item For the second baseline problem, we consider joint optimization of EV charging and V2G. In addition to the EV charging similar to the first baseline, there is EV discharge to perform V2G, i.e., $p_{dis}^{i,t} = p_{V2G}^{i,t}$. The objective function for baseline only considering V2G given by,
\begin{equation}
\begin{aligned}
& \min \quad && C_{total} = \sum_{i \in \mathcal{N}} (C_{grid}^{i}+ C_{battery}^{i}- C_{V2G}^{i})\\
& \text{subject to} && \text{Constraints}~ \eqref{cons:dynamics}-\eqref{cons:common_constraints}.
\label{obj:baseline2}
\end{aligned}
\end{equation}
\end{itemize}
Comparing our proposed V2V to the above two baseline problems, we will compare the value of V2V in EV charging with and without V2G. 

\section{Experiments and Results}\label{sec:simulation}
\subsection{Simulation Setup}
\begin{figure}[!t] 
  \centering 
    \includegraphics[width=0.9\linewidth]{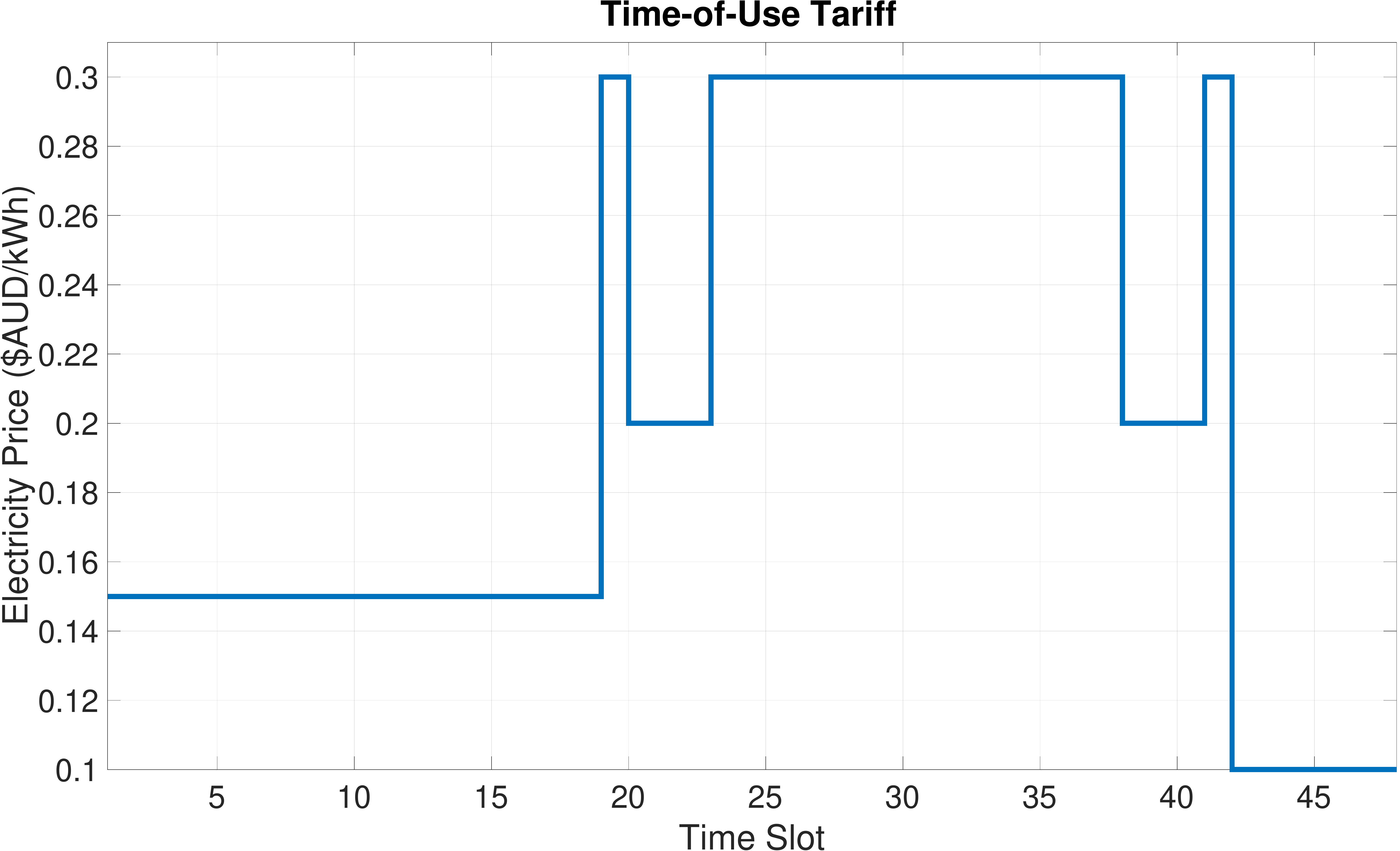} 
  \caption{Time-of-Use Tariff in 48 time slots.} 
  \label{fig:TOU tariff}  
\end{figure}

\begin{figure}[!t] 
  \centering 
    \includegraphics[width=1\linewidth]{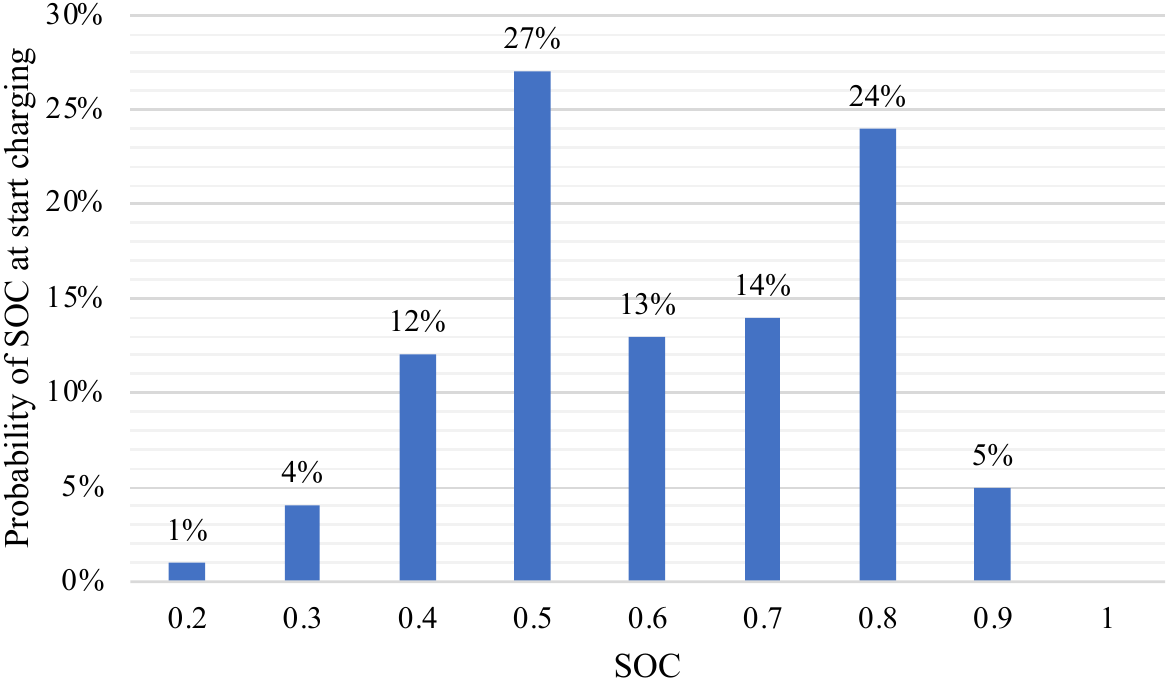} 
  \caption{Probability of EVs’ SOC at the arrival time \cite{leou2015optimal}.} 
  \label{fig:EV arrival SOC}  
\end{figure}

We consider an EV parking space with chargers (e.g., an EV charging station), where all parked EVs can perform bidirectional energy transactions. The maximum charging rate and battery capacity of all EVs are set to be $7$kW and $50$kWh, respectively. There are two different cases for EV parking space: a residential charging station in Case 1 and a commercial charging station in Case 2. Specifically, in Case 1, EVs are in a residential parking space with an operational horizon of a day (24 hours), where every 30 minutes is a time slot. For the simulation of case 1, we settled 50 EVs parked at the charging station for 24 hours, indicating the EV owners are at home. Another 50 EVs arrive and depart randomly according to a Gaussian distribution \cite{lee2011stochastic}. In Case 2, we consider a shopping center parking space with an operational horizon of 9 hours, from 9 AM to 6 PM as its business hour. In this case, 30 EVs are at a parking space for the entire 9 hours to simulate staff parking. In addition, we randomly set the arriving and departure times for another 70 EVs, simulating customer parking. Fig.~\ref{fig:TOU tariff} presents the time-of-use (TOU) tariff for EV charging, where the unit of the electricity price is the Australian dollar (\$AUD/kWh), and the unit of cost is \$AUD. The initial stored energy of EVs is illustrated in Fig.~\ref{fig:EV arrival SOC} and desired energy at departure time is distributed from 70\% to 100\% of the battery capacity. In addition, inspired by \cite{nguyen2020joint}, we introduce Jain's fairness index (JFI) to measure the fairness of energy trading among EVs defined as
\begin{equation}
JFI = \frac{(\sum_{i,j \in \mathcal{N}}\overline{P}_{V2V}^{i,j})^2}{R\sum_{i,j \in \mathcal{N}}(\overline{P}_{V2V}^{i,j})^2},
\label{eq:22}
\end{equation}
where $\overline{P}_{V2V}^{i,j}$ represents the average amount of V2V energy trading in 48 time slots for EV $i$ and $R$ is the number of EVs discharge to perform V2V. Note that the higher the JFI value is, the fairer the EV discharge is.

\subsection{Simulation Results and Analysis}
\subsubsection{Case 1: Simulation results for residential car parking space} 
\begin{figure}[!t] 
  \centering 
    \includegraphics[width=0.9\linewidth]{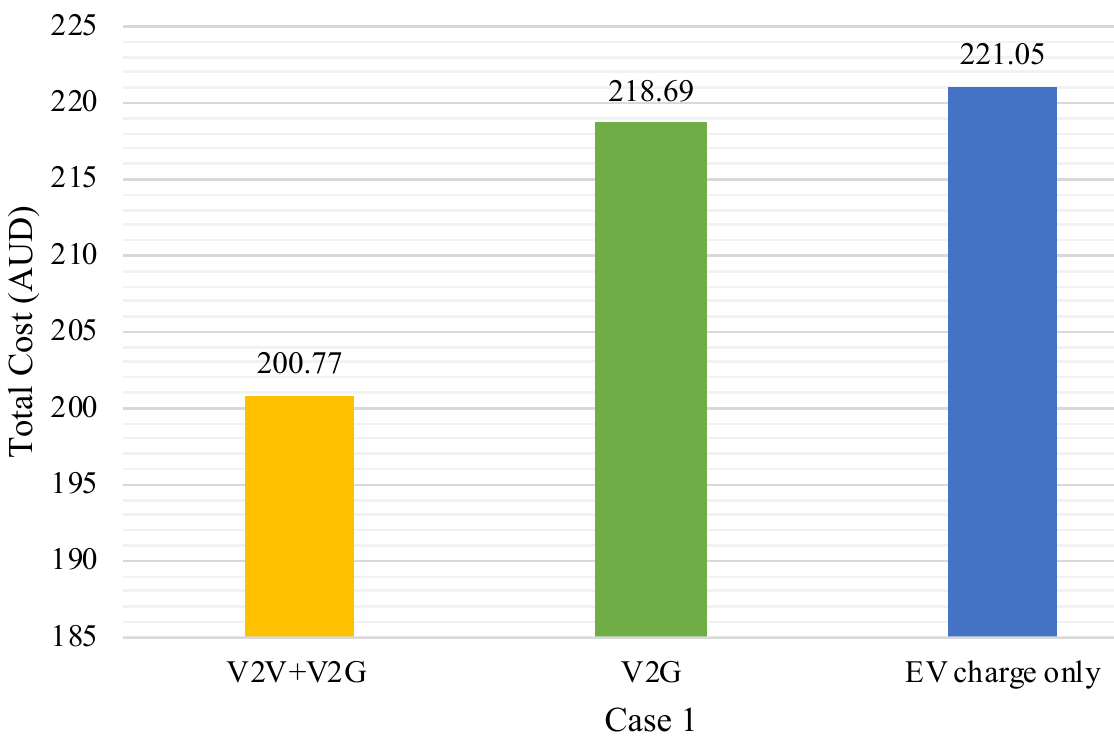} 
  \caption{Comparison of the cost of three scenarios in Case 1.} 
  \label{fig:Cost in case 1}  
\end{figure}

In this case, we consider 24 hours horizon (48-time slots) in the simulation. The EVs' traffic flow and availability in the parking space are considered for an entire day (24 hours). Fig.~\ref{fig:Cost in case 1} illustrates the minimized energy transfer cost for charging 100 vehicles in a residential parking space, where we consider three different scenarios, namely V2V+V2G joint model, V2G, and EV charging only, respectively. The vertical axis represents the total cost in Australian dollars for charging all parked EVs to predefined target values under three scenarios. The model operated on the V2G approach achieved nearly 1\% cost reduction compared to the baseline with EV charging only. The joint model (V2V + V2G) achieved the highest cost reduction of nearly 9.17\% compared to the baseline. This demonstrates the effectiveness of the joint model (V2V+V2G) in achieving a higher cost reduction than V2G only. 

\begin{figure}[!t] 
  \centering 
    \includegraphics[width=0.9\linewidth]{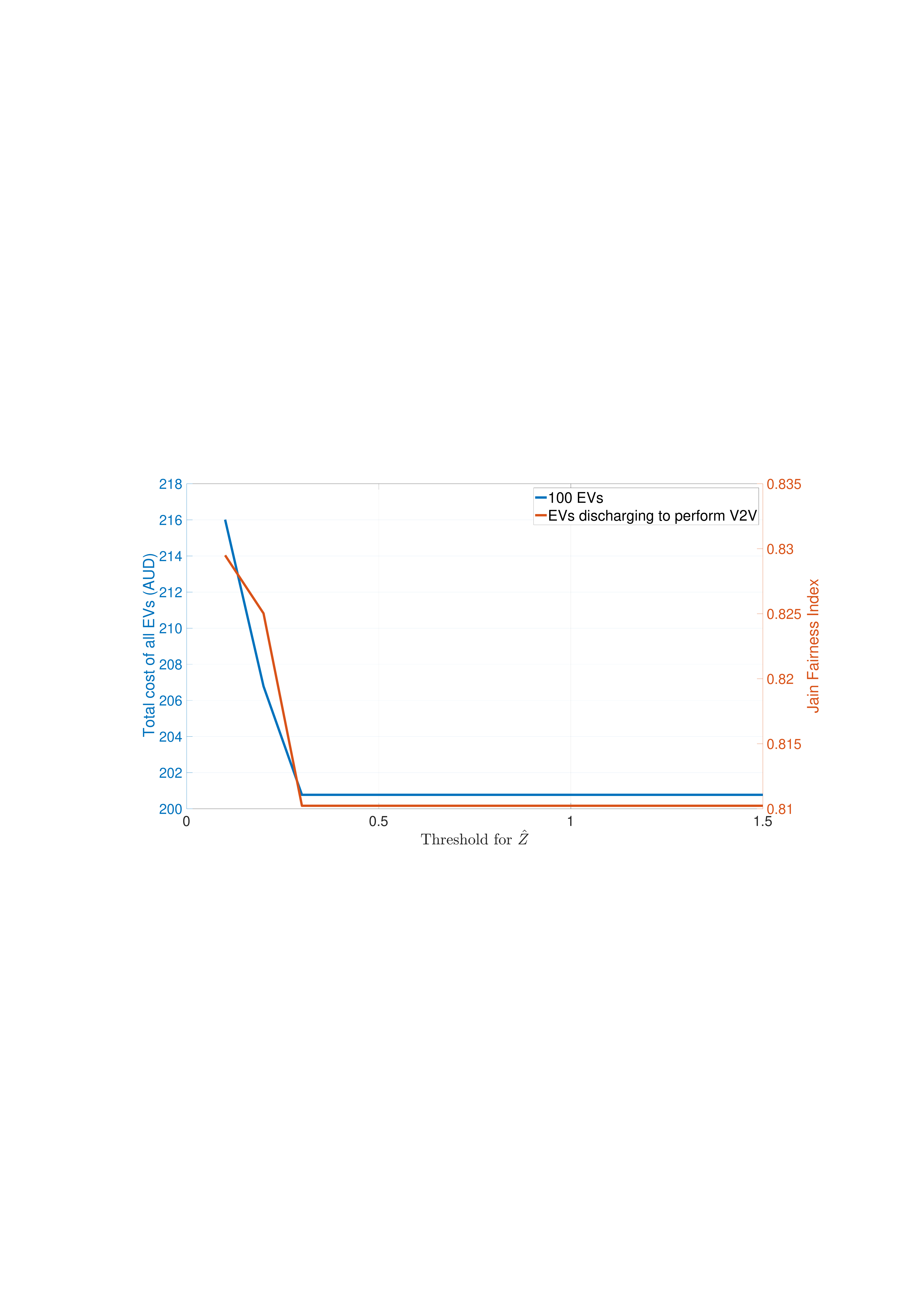} 
  \caption{Total cost and Jain fairness index for V2V+V2G model based on fairness definite 1 in Case 1.} 
  \label{fig:V2V F1 C1}  
\end{figure}

\begin{figure}[!t] 
  \centering 
    \includegraphics[width=0.9\linewidth]{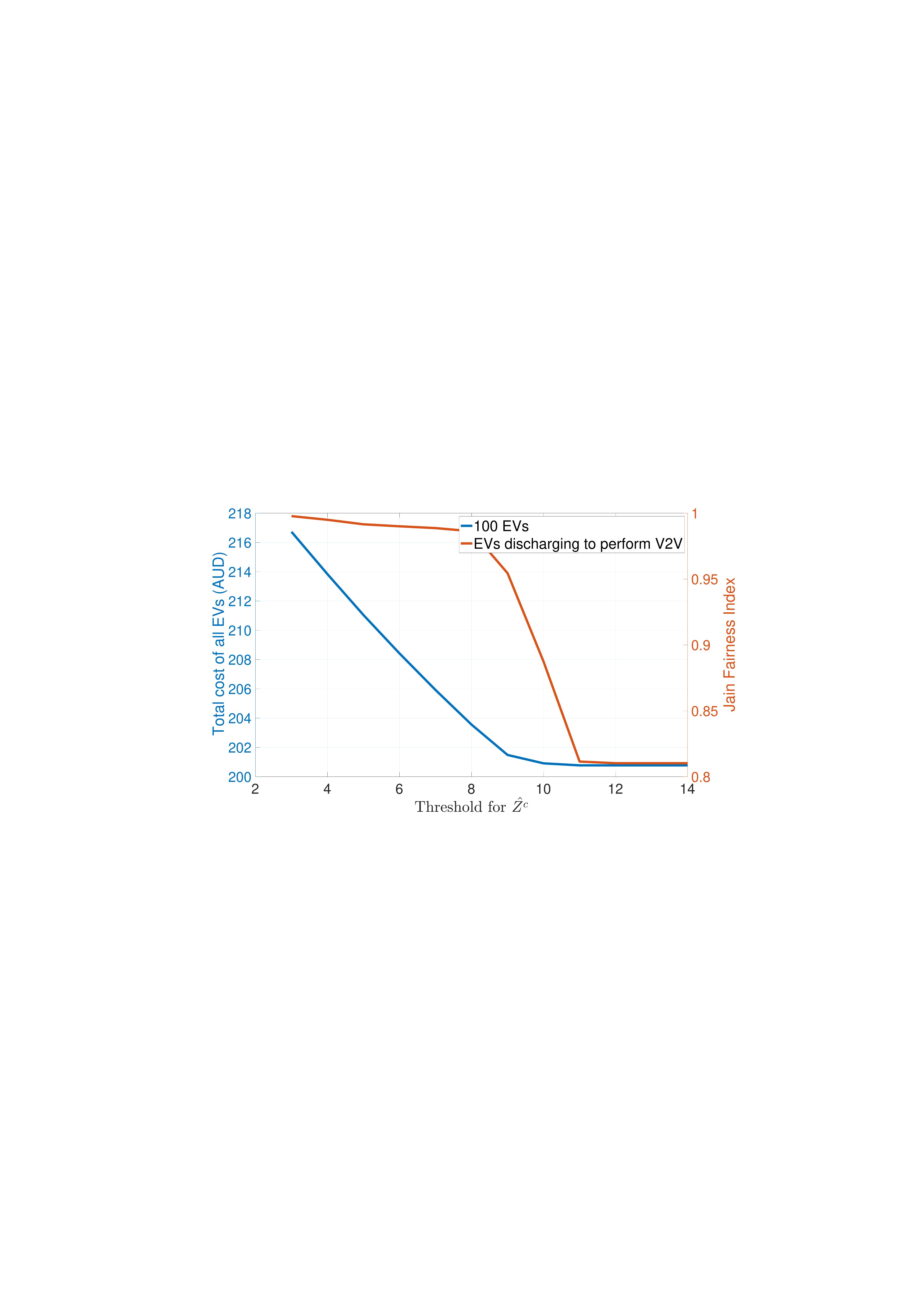} 
  \caption{Total cost and Jain fairness index for V2V+V2G model based on fairness definite 2 in Case 1.} 
  \label{fig:V2V F2 C1}  
\end{figure}

\begin{figure}[!t] 
  \centering 
    \includegraphics[width=0.9\linewidth]{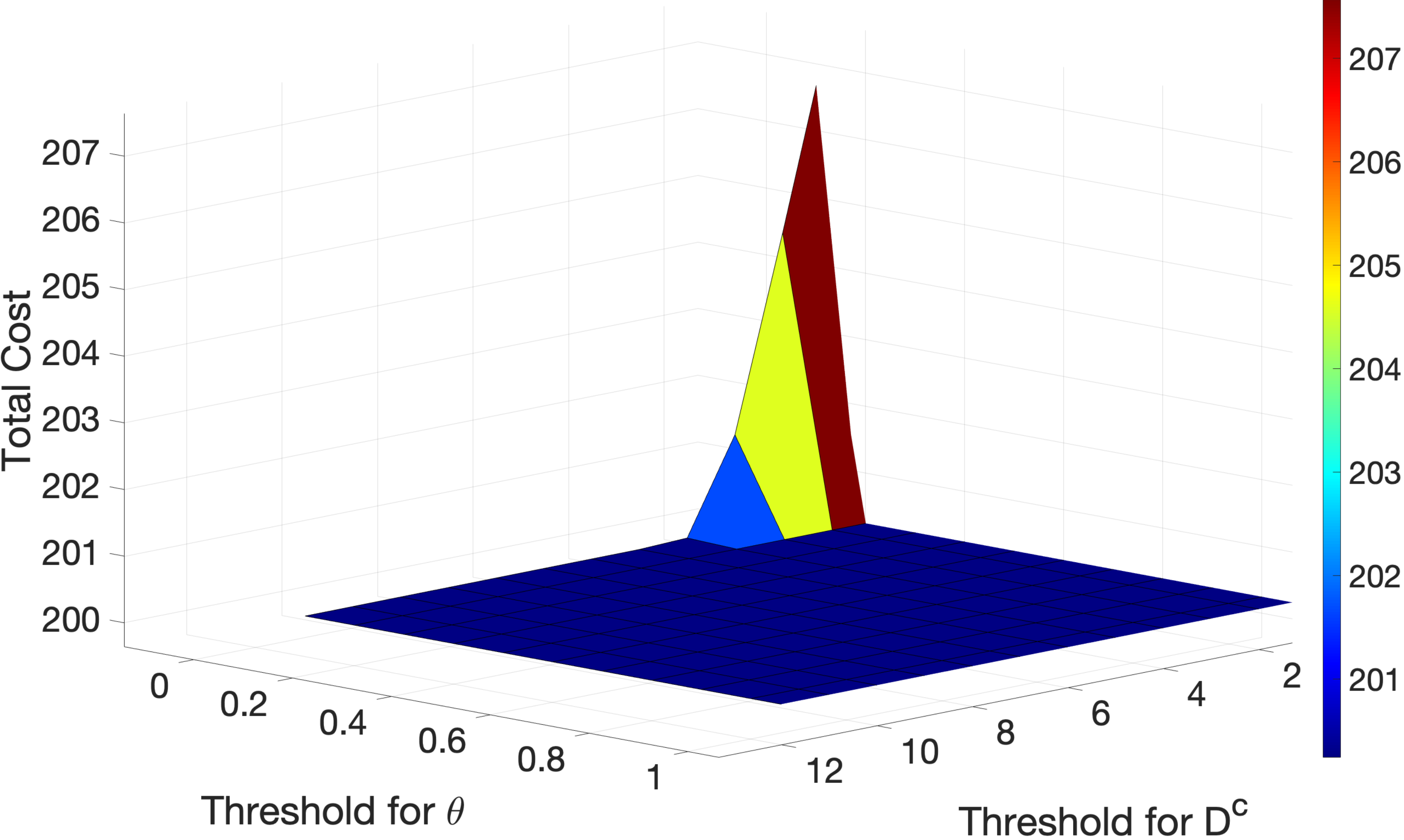} 
  \caption{Total cost for V2V+V2G model based on fairness definite 3 in Case 1.} 
  \label{fig:F3 C1 1}  
\end{figure}

\begin{figure}[!t] 
  \centering 
    \includegraphics[width=0.9\linewidth]{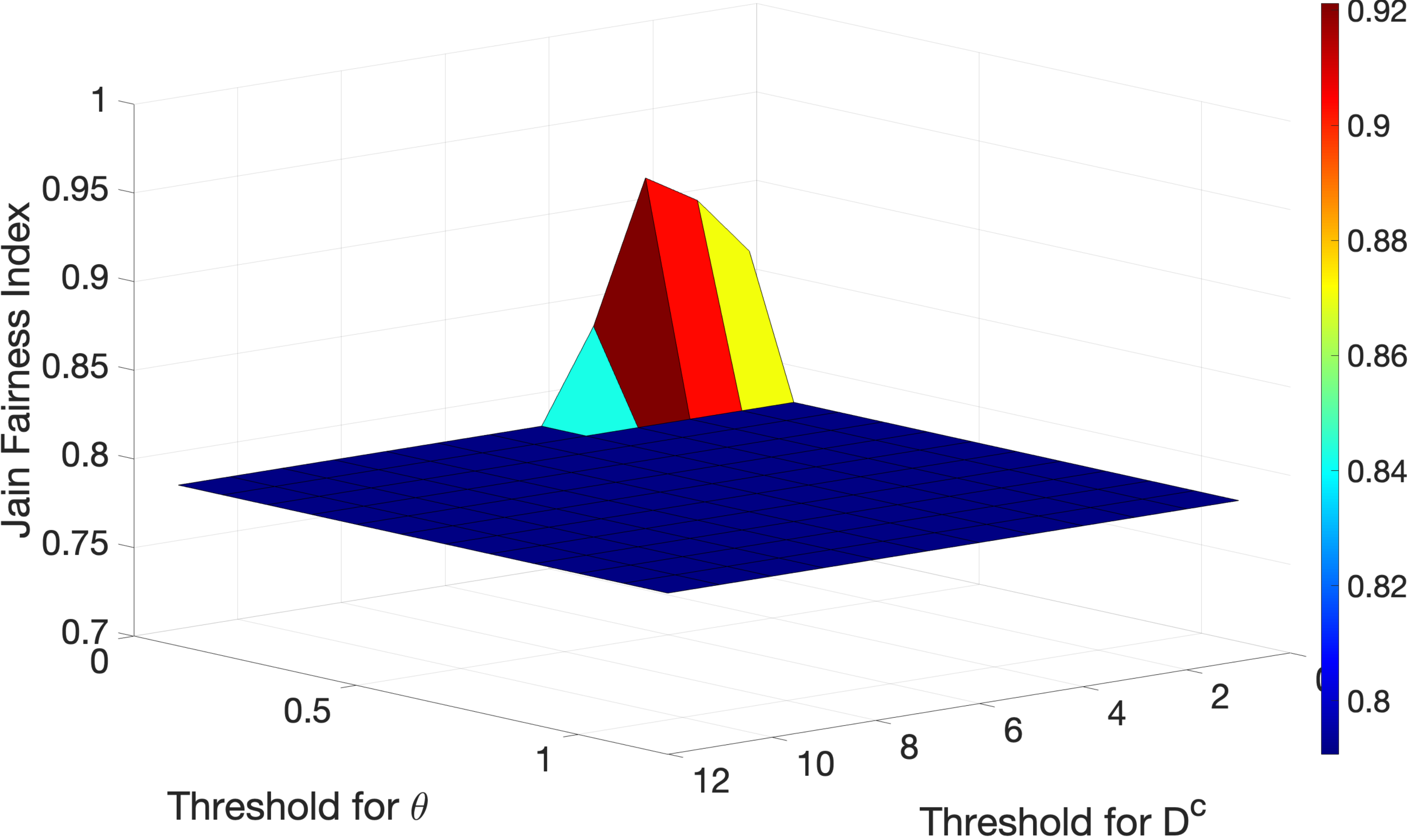} 
  \caption{Jain fairness index for V2V+V2G model based on fairness definite 3 in Case 1.} 
  \label{fig:F3 C1 2}  
\end{figure}

Then we evaluate the impact of fairness constraints and the tradeoff between fairness and EV charging cost.
Fig.~\ref{fig:V2V F1 C1} illustrates the change in the total cost and Jain fairness index based on the joint model considering the fairness constraints with fix limit. We can see that the value of threshold $\overline{\mathcal{Z}}$ is in the range of 0.1 to 1.5. The total cost of the joint model varies from 200.77 to 216.01 and the Jain fairness index from 0.81 to 0.83. That means the higher value of threshold $\overline{\mathcal{Z}}$ is, the lower the total cost is. The results of the joint model based on the second fairness constraint are presented in Fig.~\ref{fig:V2V F2 C1}, which are similar to the results compared to the first fairness constraint. The threshold $\overline{\mathcal{Z}^c}$ ranges from 1 to 14 and we can see that the Jain fairness index in the joint model is more sensitive, since the range of the Jain fairness index in fairness two is from 0.81 to 0.98. Fig.~\ref{fig:F3 C1 1} and Fig.~\ref{fig:F3 C1 2} show the results considering the third fairness constraint with a budget. We can see that both thresholds $\theta$ and $\mathcal{D}_{i}^c$ affect the total cost and Jain fairness index of the joint model, where the higher $\theta$ and $\mathcal{D}_{i}^c$ are, the lower total cost and the V2V energy trading fairness become.

\subsubsection{Case 2: Simulation results for shopping center car parking space}
\begin{figure}[!t] 
  \centering 
    \includegraphics[width=0.9\linewidth]{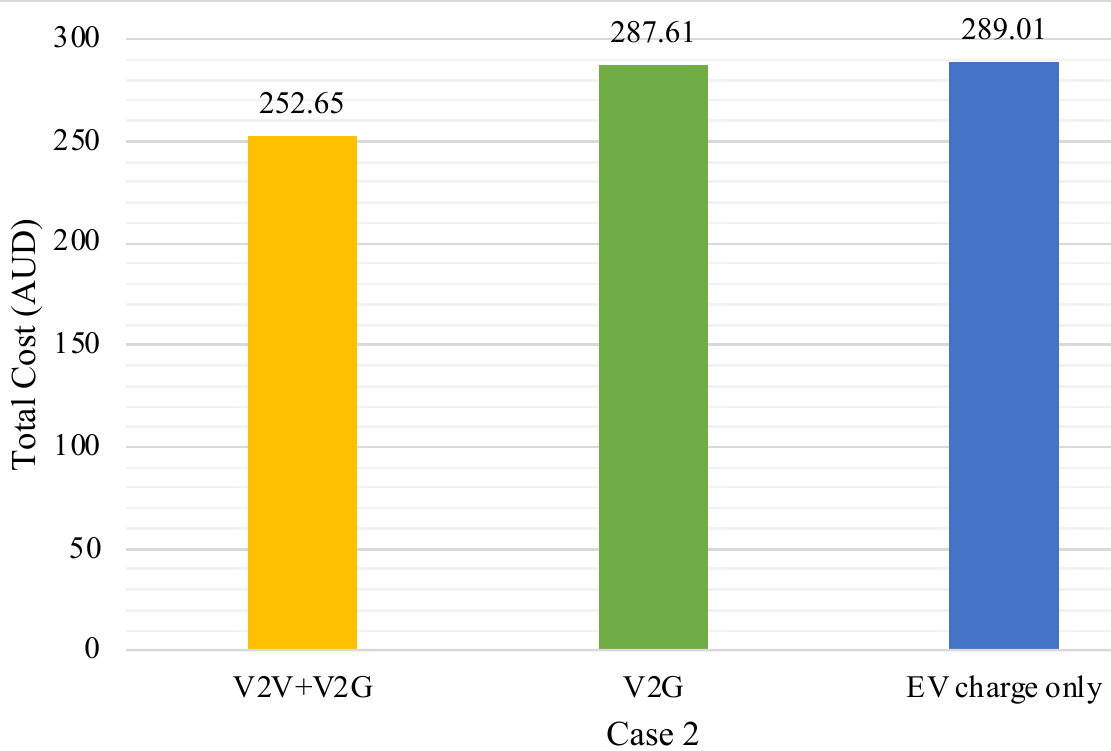} 
  \caption{Comparison of the cost of three scenarios in Case 2.} 
  \label{fig:cost in case 2}  
\end{figure}

Fig.~\ref{fig:cost in case 2} shows the minimized energy cost for charging 100 vehicles parked in the shopping center parking space. Similar to the results of case 1, the joint model achieves the highest cost reduction. It costs \$AUD 289.01 to charge the parked vehicles to their specified target values by only using energy from the grid. While V2G achieves an insignificant cost reduction of 1\%. The joint model (V2G+V2V) is proven to be the most cost-effective, with a cost reduction of 12.58\% in the case of the shopping center parking.

\begin{figure}[!t] 
  \centering 
    \includegraphics[width=0.9\linewidth]{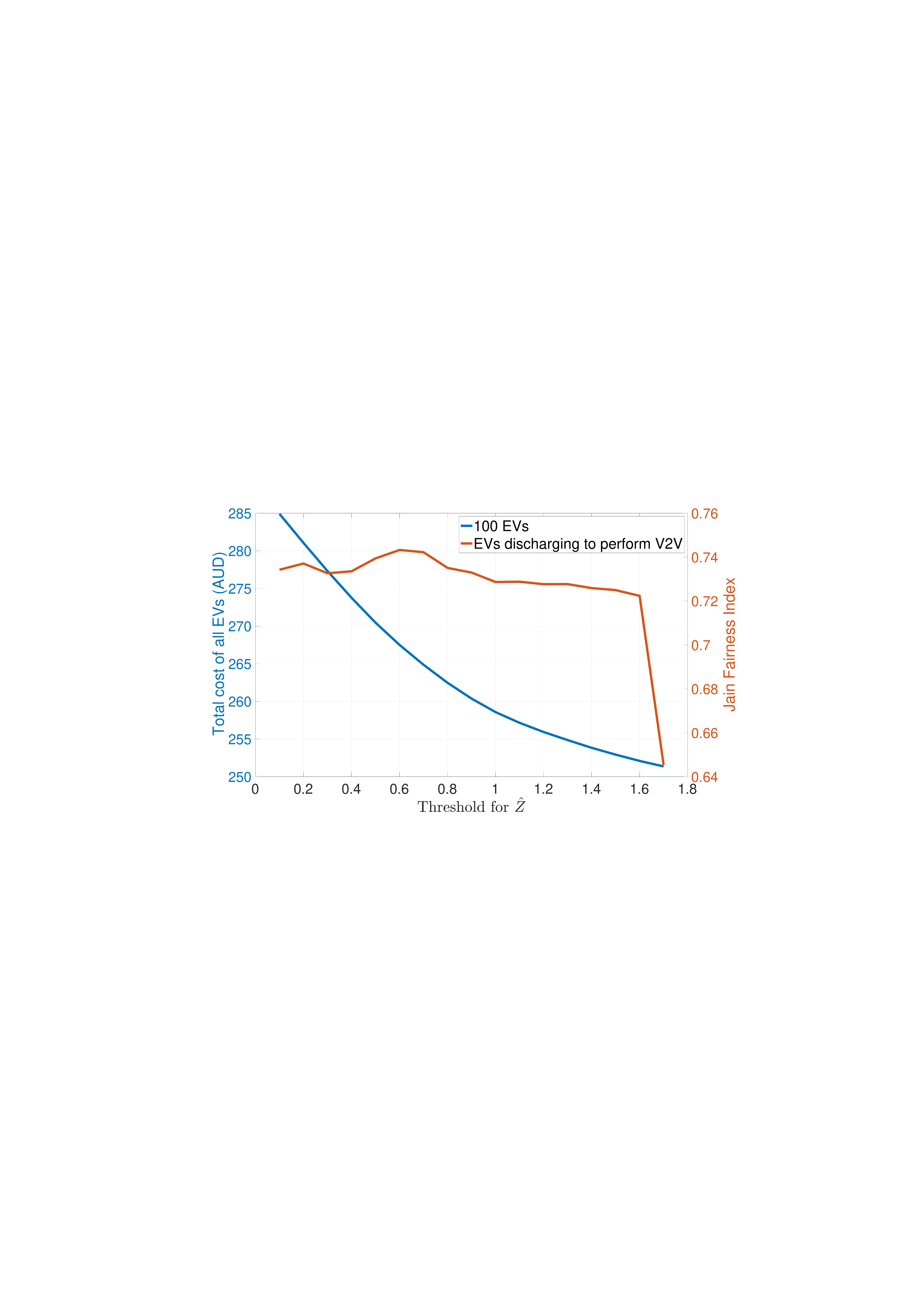} 
  \caption{Total cost and Jain fairness index for V2V+V2G model based on fairness definite 1 in Case 2.} 
  \label{fig:V2V F1 C2}  
\end{figure}

\begin{figure}[!t] 
  \centering 
    \includegraphics[width=0.9\linewidth]{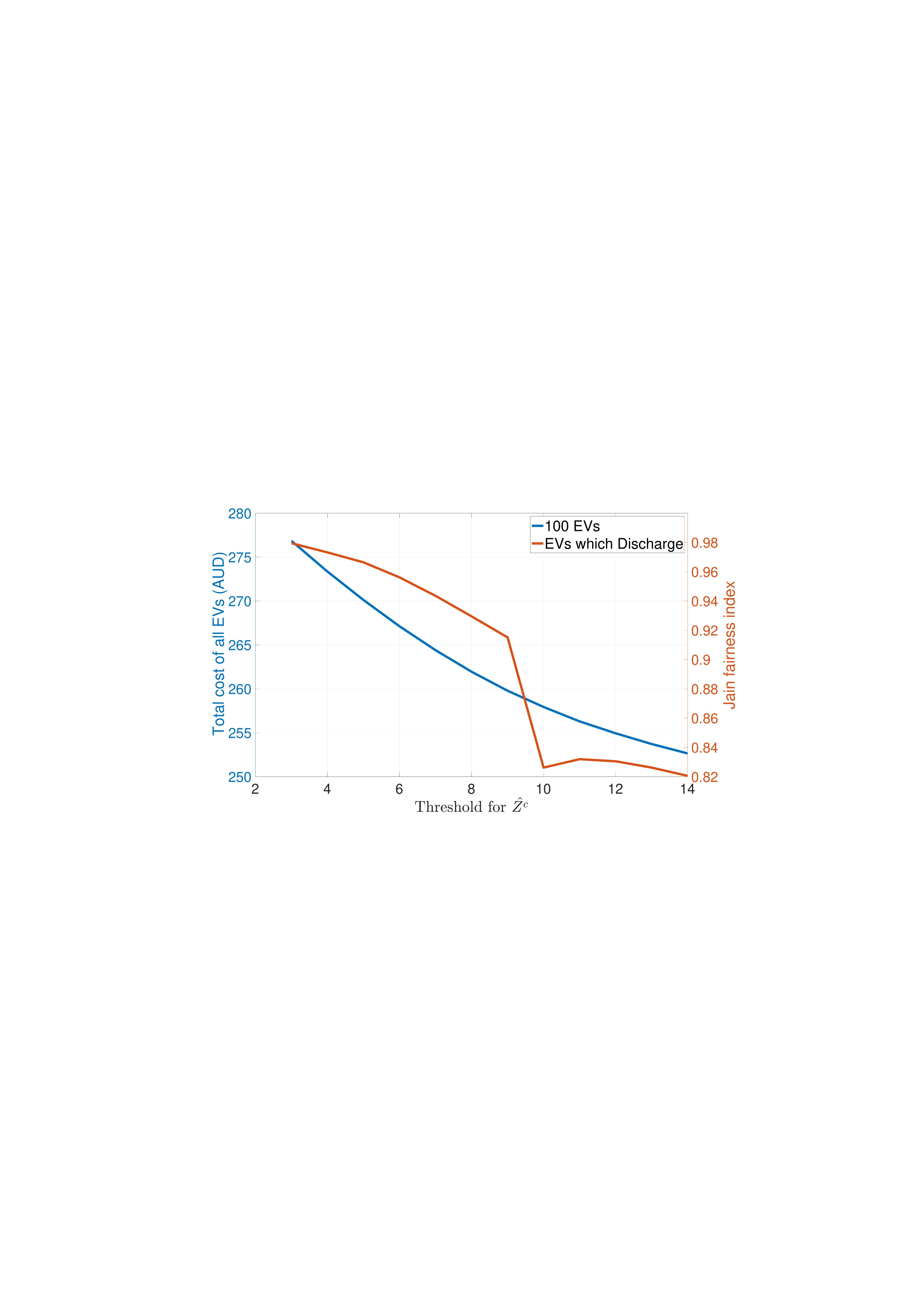} 
  \caption{Total cost and Jain fairness index for V2V+V2G model based on fairness definite 2 in Case 2.} 
  \label{fig:V2V F2 C2}  
\end{figure}

\begin{figure}[!t] 
  \centering 
    \includegraphics[width=0.9\linewidth]{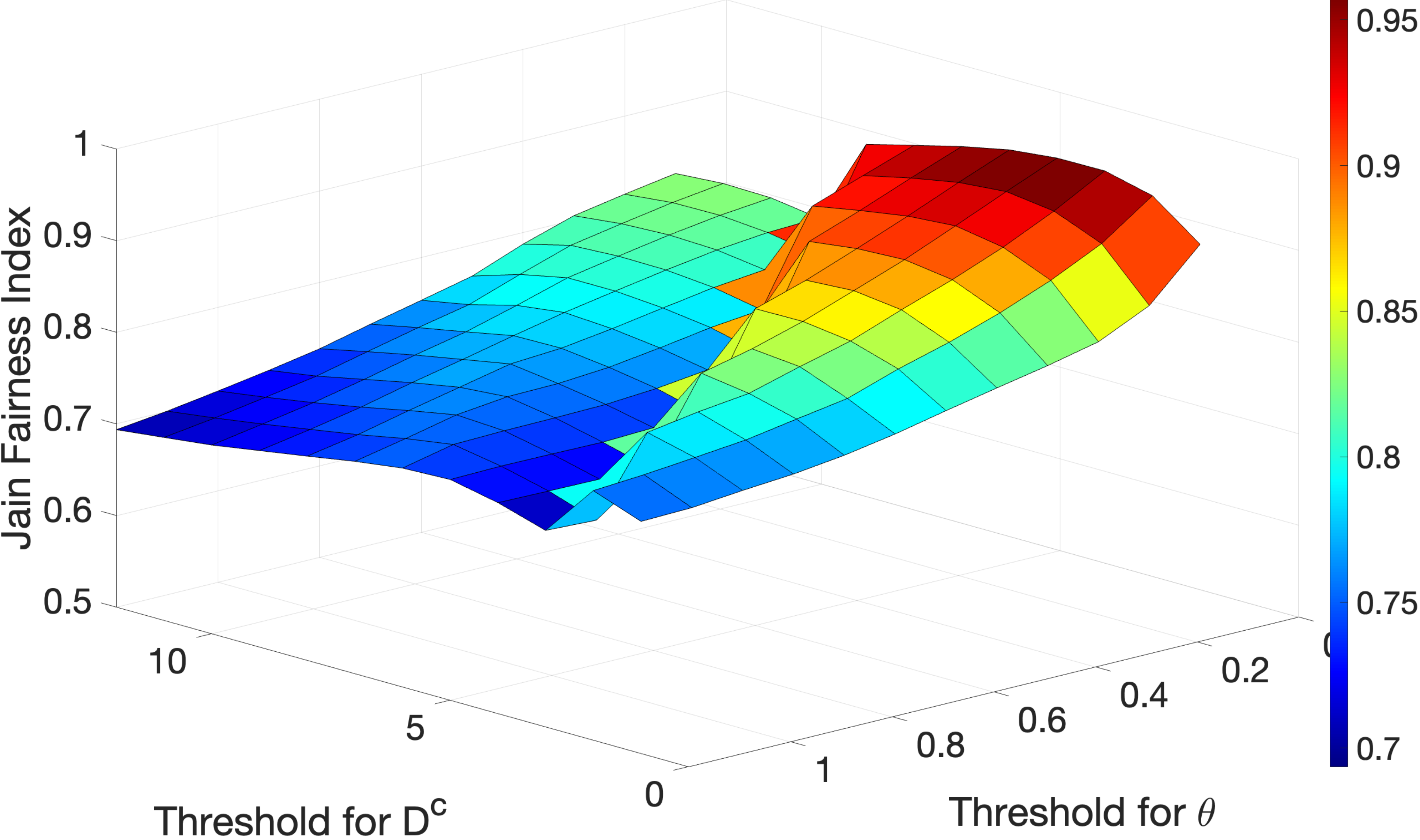} 
  \caption{Jain fairness index for V2V+V2G model 1 based on fairness definite 3 in Case 2.} 
  \label{fig:F3 C2 1}  
\end{figure}

\begin{figure}[!t] 
  \centering 
    \includegraphics[width=0.9\linewidth]{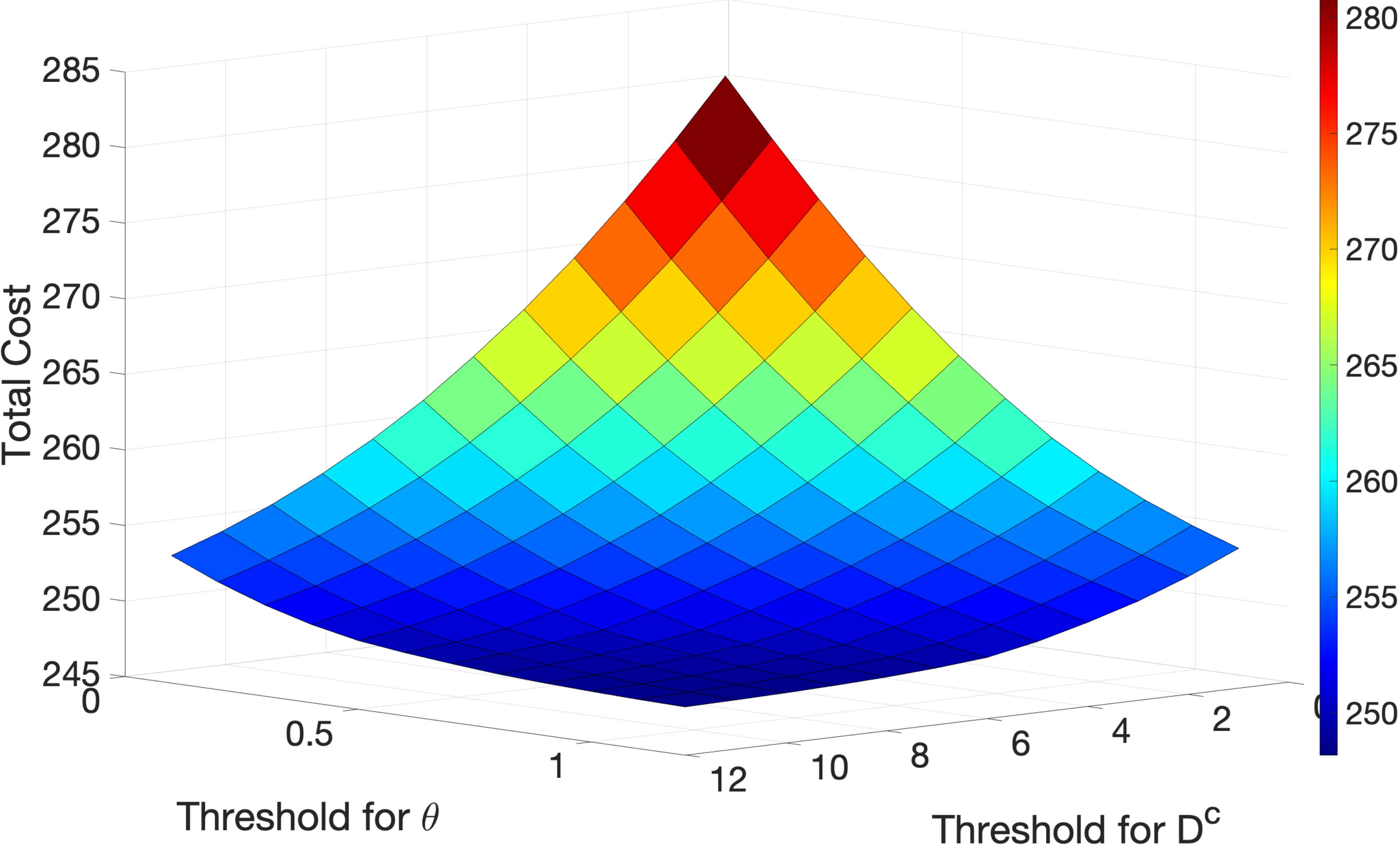} 
  \caption{Total cost for V2V+V2G model 1 based on fairness definite 3 in Case 2.} 
  \label{fig:F3 C2 2}  
\end{figure}

Fig.~\ref{fig:V2V F1 C2} presents the change of total cost and Jain fairness index based on the V2V+V2G model considering the fairness constraints with fixed limit. We can see that the higher value of threshold $\overline{\mathcal{Z}}$ is, the lower the total cost is, where the total cost ranges from 252.65 to 285.01. However, with the increase of fairness threshold $\overline{\mathcal{Z}}$, the value of Jain fairness index fluctuates slightly from 0.73 to 0.74 when $\overline{\mathcal{Z}}$ range from 0.1 to 0.7. 
The results based on the second fairness constraint are presented in Fig.~\ref{fig:V2V F2 C2}, which are similar to the results compared to the first fairness constraint. With the increase of threshold $\overline{\mathcal{Z}^c}$, the total cost of the joint model increases from 252.65 to 276.02. 
Fig.~\ref{fig:F3 C2 1} and Fig.~\ref{fig:F3 C2 2} show the results considering the third fairness constraint with a budget for the joint model. Similarly, we can see both thresholds $\theta$ and $\mathcal{D}_{i}^c$ affect the total cost and Jain fairness index.  The total cost and the EV discharging fairness decrease with an increasing $\theta$ and $\mathcal{D}_{i}^c$.

\section{Conclusion}\label{sec:conclusion}
In this paper, we proposed fairness-aware optimization for the joint coordination of EV charging, V2G, and V2V, while preventing unbiased discharging of EVs. We considered two discharging options for EVs, which are V2G and V2V. Furthermore, we defined fairness metrics for V2V activity to avoid excessive discharge of EVs and included fairness constraints in the optimization problem to maintain the fair contribution of EVs. We used the time-of-use (TOU) scheme for energy purchase prices and wholesale market prices for selling energy back to the grid. The resulting costs of charging EVs to the desired levels by different methods were compared. The numerical results showed the model with the combination of V2G and V2V systems achieves better performance than V2G and only energy charging alone, with a cost reduction of 9.17\% in the case of residential car park whereas 12.58\% in the case of the shopping car park. In addition, the simulation results showed a tradeoff between economic performance, charging cost, and fairness of discharging among EVs.

\bibliographystyle{IEEEtran}
\bibliography{ref.bib}

\end{document}